\documentclass[10pt,english]{article}
\usepackage[T1]{fontenc}
\usepackage[latin9]{inputenc}
\usepackage{geometry}
\usepackage{color}
\usepackage{graphicx}

\makeatletter
\usepackage{cite}

\makeatother

\usepackage{babel}
\begin{document}

\title{Comment on ``Quantum Teleportation of Eight-Qubit State via Six-Qubit
Cluster State''}

\author{Mitali Sisodia and Anirban Pathak}
\maketitle
\begin{center}
Jaypee Institute of Information Technology, A 10, Sector 62, Noida,
UP 201309
\par\end{center}
\begin{abstract}
Recently, Zhao et al., (Int. J. Theor. Phys. \textbf{57},\textbf{
}516\textendash 522 (2018)) have proposed a scheme for quantum teleportation
of an eight-qubit quantum state using a six qubit cluster state. In
this comment, it's shown that the quantum resource (multi-partite
entangled state used as the quantum channel) used by Zhao et al.,
is excessively high and the task can be performed\textcolor{red}{{}
}using any two Bell states as the task can be reduced to the teleportation
of an arbitrary two qubit state. Further, a trivial conceptual mistake
made by Zhao et al., in the description of the quantum channel has
been pointed out. It's also mentioned that recently a trend of proposing
teleportation schemes with excessively high quantum resources has
been observed and the essence of this comment is applicable to all
such proposals.
\end{abstract}

\section{Introduction}

The original protocol for quantum teleportation was designed for the
teleportation of a single qubit state using a Bell state \cite{ben}.
Subsequently, many schemes have been proposed for the teleportation
of multi-qubit states using various entangled states. Following the
trend, recently, Zhao et al., have proposed a scheme for the teleportation
of the following quantum state

\begin{equation}
\begin{array}{lcc}
|\varphi\rangle_{abcdefgh} & = & \left(\alpha|00000000\rangle+\beta|00100000\rangle+\gamma|11011111\rangle+\delta|11111111\rangle\right)_{abcdefgh}\end{array},\label{eq:1}
\end{equation}
where the coefficients $\alpha,\,\beta,\,\gamma,\,\delta$ are unknown
and satisfies $\left|\alpha\right|^{2}+\left|\beta\right|^{2}+\left|\gamma\right|^{2}+\left|\delta\right|^{2}=1$
(cf. Eq. (1) of \cite{zhao}). They considered this state as a 8-qubit
quantum state and proposed a scheme for teleportation of this state
using a six-qubit cluster state described as

\begin{equation}
\begin{array}{lcc}
|\phi\rangle_{123456} & = & \left(\alpha|000000\rangle+\beta|001001\rangle+\gamma|110110\rangle+\delta|111111\rangle\right)_{123456}\end{array}\label{eq:2-1}
\end{equation}
as the quantum channel (cf. Eq. (2) of \cite{zhao}). The basic conceptual
problem with the form of this channel is that this channel cannot
be constituted as the coefficients $\alpha,\,\beta,\,\gamma,\,\delta$
present in the expansion of $|\varphi\rangle_{abcdefgh}$ are unknown.
Further, Eq. (2) of \cite{zhao} is not consistent with Eqs. (5)-(6)
of \cite{zhao} and thus with the remaining part of \cite{zhao}.
To stress on the more important issues and to continue the discussion,
we may consider that Zhao et al., actually intended to use 
\begin{equation}
|\phi\rangle_{123456}=\frac{1}{2}\left(|000000\rangle+|001001\rangle+|110110\rangle+|111111\rangle\right),\label{eq:tin}
\end{equation}
which is consistent with Eq. (3) of \cite{zhao}. However, the above
mistake does not appear to be a typographical error as the same error
is present in another recent work of the authors (see \cite{chen}).The
more important question is whether we need Eq. (\ref{eq:tin}) for
the teleportation of $|\varphi\rangle_{abcdefgh}$, or the task can
be performed using a\textcolor{red}{{} }simpler quantum channel. This
is the question, we wish to address in this comment. Here, it is important
to note that in a recent work \cite{sisodia1}, we have shown that
a quantum state having $n$ unknown coefficients can be teleported
by using $\log_{2}\left\lceil \frac{n}{2}\right\rceil $ number of
Bell states. Now, as there are 4 unknowns in $|\varphi\rangle_{abcdefgh},$
teleportation of this state should require only 2 Bell states. This
point can be further illustrated by noting that Zhao et al., have
shown that using 4 CNOT gates (cf. Fig. 1 of \cite{zhao}) $|\varphi\rangle_{abcdefgh}$
can be transformed to a state $|\varphi^{\prime}\rangle_{abcdefgh}=|\chi\rangle_{abcd}|0000\rangle_{efgh}$,
where 

\begin{equation}
\begin{array}{lcc}
|\chi\rangle_{abcd} & = & \left(\alpha|0000\rangle+\beta|0010\rangle+\gamma|1101\rangle+\delta|1111\rangle\right)_{abcd}\end{array}\label{eq:3}
\end{equation}
and thus the actual teleportation task reduces to the teleportation
of $|\chi\rangle_{abcd}$. Now, we can introduce a circuit which can
be described as ${\rm CNOT}_{a\rightarrow d}{\rm {\rm CNOT}_{a\rightarrow b}SWAP}_{bc}$,
where ${\rm SWAP}_{ij}$ and ${\rm CNOT}_{i\rightarrow j}$ correspond
to a gate that swaps $i$th and $j$th qubit and a ${\rm CNOT}$ that
uses $i$th qubit as the control qubit and $j$th qubit as the target
qubit, respectively. On application of this circuit, $|\chi\rangle_{abcd}$
would transform to $\left(\alpha|00\rangle+\beta|01\rangle+\gamma|10\rangle+\delta|11\rangle\right)_{ac}|00\rangle_{bd}$
as 
\begin{equation}
{\rm CNOT}_{a\rightarrow d}{\rm {\rm CNOT}_{a\rightarrow b}SWAP}_{bc}|\chi\rangle_{abcd}=\left(\alpha|00\rangle+\beta|01\rangle+\gamma|10\rangle+\delta|11\rangle\right)_{ac}|00\rangle_{bd}.\label{eq:new1}
\end{equation}
Thus, a scheme for teleportation of an arbitrary two qubit state $|\psi\rangle_{ac}=\left(\alpha|00\rangle+\beta|01\rangle+\gamma|10\rangle+\delta|11\rangle\right)_{ac}$
will be sufficient for the teleportation of $|\varphi\rangle_{abcdefgh}.$
Such a scheme for the teleportation of an arbitrary two qubit state
was proposed by Rigolin in 2005 \cite{Rigolin-PRA}. He proposed 16
quantum channels that are capable of teleportation of the arbitrary
2 qubit states and referred to those channels as generalized Bell
states. However, soon after the work of Rigolin, in a comment on it,
Deng \cite{PRA-comment} had shown that 16 channels introduced by
Rigolin are simply the 16 possible product states of the 4 Bell states,
and thus the work of Deng and Rigolin established the fact that the
product of any two Bell states is sufficient for the teleportation
of an arbitrary two qubit state. Result of Deng and Rigolin is consistent
with the more general recent result of ours \cite{sisodia1} and that
of \cite{thapliyal,verma,anindita}. This clearly, establishes that
in addition to the circuit block comprise of 4 CNOT gates used in
the work of Zhao et al., if one uses the circuit described above and
well known scheme of Rigolin, then one will be able to teleport so-called
8 qubit state $|\varphi\rangle_{abcdefgh}$ using only 2 Bell states.
This comment could have been concluded at this point, but for the
convenience of the readers we have added the following section, where
we briefly outline the complete process of teleportation of $|\varphi\rangle_{abcdefgh}$
using 2 Bell states.

\begin{figure}
\begin{centering}
\includegraphics[scale=0.65]{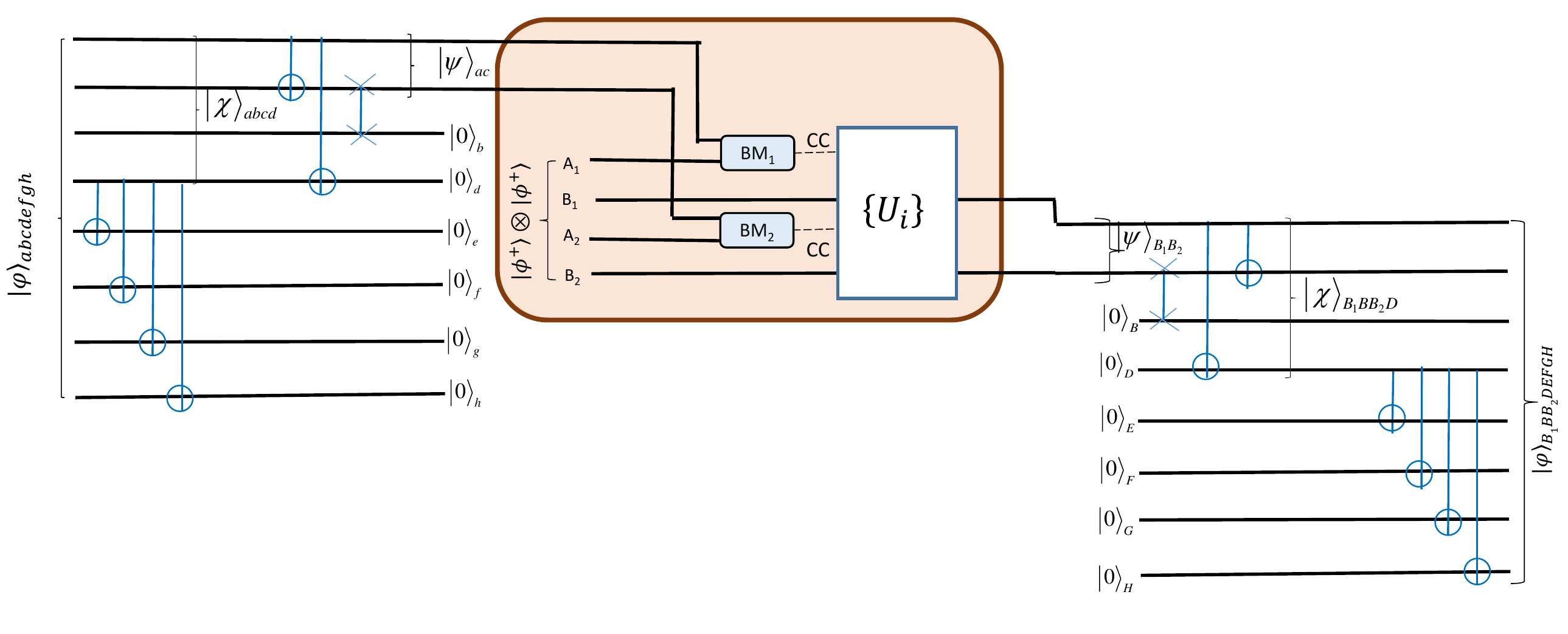}
\par\end{centering}
\caption{\label{fig:ckt}(Color online) Circuit for teleportation of the quantum
state used in Ref. \cite{zhao} (described by Eq. (\ref{eq:1}) of
this paper) using the optimal quantum resource, i.e., two Bell states.
BM corresponds to the Bell measurement and CC corresponds to the classical
communication of the measurement outcome from Alice to Bob, which
enables Bob to apply unitary operation $U_{i}$. The left most block
of the circuit is used to reduce the task of teleporting an 8 qubit
state to that of a 2 qubit state. The middle block describes a scheme
for teleportation of an arbitrary 2 qubit states using 2 Bell states,
and the right most circuit block is used here to locally transform
the teleported 2 qubit state in to the desired 8 qubit state which
we originally aim to teleport.}
\end{figure}

\section{Complete teleportation process}

Teleportation of $|\varphi\rangle_{abcdefgh}$ is accomplished in
3 steps as illustrated in Fig. \ref{fig:ckt}. Firstly, the task of
teleportation of the so called eight qubit quantum state $|\varphi\rangle_{abcdefgh}$
is reduced to the task of teleporting a two qubit quantum state $|\psi\rangle_{ac}$
by transforming $|\varphi\rangle_{abcdefgh}$ into $|\psi\rangle_{ac}|000000\rangle_{bdefgh}$.
This is done using the circuit block (comprised of 6 CNOT gates and
1 SWAP gate) shown in the left side of the complete circuit shown
in Fig. \ref{fig:ckt}. The actual teleportation part is the second
step which is shown in the middle block of Fig. \ref{fig:ckt} (cf.
the rectangular box in the middle of the Fig. \ref{fig:ckt}). In
this step we teleport $|\psi\rangle_{ac}$ using two Bell states as
quantum channel. As an example, consider . $|\phi^{+}\rangle_{A_{1}B_{1}}\otimes|\phi^{+}\rangle_{A_{2}B_{2}}=$$\frac{|00\rangle+|11\rangle}{\sqrt{2}}\otimes\frac{|00\rangle+|11\rangle}{\sqrt{2}}$
as the quantum channel. Thus, the combined state would be 

\begin{equation}
\begin{array}{lcc}
|\Phi\rangle_{acA_{1}B_{1}A_{2}B_{2}} & = & |\psi\rangle_{ac}\otimes|\phi^{+}\rangle_{A_{1}B_{1}}\otimes|\phi^{+}\rangle_{A_{2}B_{2}},\end{array}\label{eq:5}
\end{equation}
which can be decomposed as 

\begin{equation}
\begin{array}{lcl}
|\Phi\rangle_{acA_{1}B_{1}A_{2}B_{2}} & = & |\phi^{+}\rangle_{aA_{1}}|\phi^{+}\rangle_{cA_{2}}I\otimes I\,|\psi\rangle_{B_{1}B_{2}}+|\phi^{+}\rangle_{aA_{1}}|\phi^{-}\rangle_{cA_{2}}I\otimes Z\,|\psi\rangle_{B_{1}B_{2}}\\
 & + & |\phi^{+}\rangle_{aA_{1}}|\psi^{+}\rangle_{cA_{2}}I\otimes X\,|\psi\rangle_{B_{1}B_{2}}+|\phi^{+}\rangle_{aA_{1}}|\psi^{-}\rangle_{cA_{2}}I\otimes iY\,|\psi\rangle_{B_{1}B_{2}}\\
 & + & |\phi^{-}\rangle_{aA_{1}}|\phi^{+}\rangle_{cA_{2}}Z\otimes I\,|\psi\rangle_{B_{1}B_{2}}+|\phi^{-}\rangle_{aA_{1}}|\phi^{-}\rangle_{cA_{2}}Z\otimes Z\,|\psi\rangle_{B_{1}B_{2}}\\
 & + & |\phi^{-}\rangle_{aA_{1}}|\psi^{+}\rangle_{cA_{2}}Z\otimes X\,|\psi\rangle_{B_{1}B_{2}}+|\phi^{-}\rangle_{aA_{1}}|\psi^{-}\rangle_{cA_{2}}Z\otimes iY\,|\psi\rangle_{B_{1}B_{2}}\\
 & + & |\psi^{+}\rangle_{aA_{1}}|\phi^{+}\rangle_{cA_{2}}X\otimes I\,|\psi\rangle_{B_{1}B_{2}}+|\psi^{+}\rangle_{aA_{1}}|\phi^{-}\rangle_{cA_{2}}X\otimes Z\,|\psi\rangle_{B_{1}B_{2}}\\
 & + & |\psi^{+}\rangle_{aA_{1}}|\psi^{+}\rangle_{cA_{2}}X\otimes X\,|\psi\rangle_{B_{1}B_{2}}+|\psi^{+}\rangle_{aA_{1}}|\psi^{-}\rangle_{cA_{2}}X\otimes iY\,|\psi\rangle_{B_{1}B_{2}}\\
 & + & |\psi^{-}\rangle_{aA_{1}}|\phi^{+}\rangle_{cA_{2}}iY\otimes I\,|\psi\rangle_{B_{1}B_{2}}+|\psi^{-}\rangle_{aA_{1}}|\phi^{-}\rangle_{cA_{2}}iY\otimes Z\,|\psi\rangle_{B_{1}B_{2}}\\
 & + & |\psi^{-}\rangle_{aA_{1}}|\psi^{+}\rangle_{cA_{2}}iY\otimes X\,|\psi\rangle_{B_{1}B_{2}}+|\psi^{-}\rangle_{aA_{1}}|\psi^{-}\rangle_{cA_{2}}iY\otimes iY\,|\psi\rangle_{B_{1}B_{2}}.
\end{array}\label{eq:6}
\end{equation}
Eq. (\ref{eq:6}) clearly shows that Alice's Bell measurement on qubits
$aA_{1}$ and $cA_{2}$ reduces Bob's qubits $B_{1}B_{2}$ in such
a quantum state that the state $|\psi\rangle_{B_{1}B_{2}}$ can be
obtained by applying local unitary operations. The choice of the unitary
operations depends on the Alice's measurement outcome (as illustrated
in Eq. (\ref{eq:6})). After teleporting $|\psi\rangle$, in the third
step, the initial state to be teleported is reconstructed from $|\psi\rangle_{B_{1}B_{2}}$
by using six ancillary qubits prepared in $\left(|0\rangle^{\otimes6}\right)_{BDEFGH}$
and the circuit block shown in the right side of Fig. \ref{fig:ckt}.
At the output of this circuit block we would obtain the so called
8 qubit state $|\varphi\rangle_{B_{1}BB_{2}DEFGH}$ that we wanted
to teleport. 

\section{Concluding remark}

Despite the existence of old results of \cite{Rigolin-PRA,thapliyal,verma,anindita,Yang-chinese-phys-lett}
and our recent results \cite{sisodia1}, people are frequently proposing
teleportation schemes \cite{chen,li,dhara} and their variants \cite{choudhury}
using higher amount of quantum resources, although they know that
the preparation and maintenance of such resources are not easy. Specially,
preparation of multi-partite entanglement is difficult. Keeping that
in mind, it is advisable that while designing new schemes, unnecessary
use of quantum resources should be circumvented. This comment is not
only applicable to the scheme of Zhao et al., the main idea of it
is applicable to many other schemes (\cite{chen,sisodia1,li,dhara,choudhury}
and references therein) as ther amount of quantum resources used by
them are higher than the minimum required amount. 

\subsection*{Acknowledgments }

AP thanks the SERB, Department of Science and Technology (DST), India,
for support provided through the project No. EMR/2015/000393. AP and
MS also thank K. Thapliyal for his interest in this work some useful
discussions.

\end{document}